\documentclass{ws-procs9x6}
\begin{document}

\title{Functionals linear in curvature and statistics of helical proteins}

\author{V.~V. NESTERENKO\footnote{\uppercase{W}ork partially
supported by grant 03-01-00025 of the \uppercase{R}ussian
\uppercase{F}oundation for \uppercase{B}asic \uppercase{R}search and by
\uppercase{ISTC} (\uppercase{P}roject  840).}}

\address{Bogoliubov Laboratory for Theoretical Physics,\\ Joint Institute
for Nuclear Research, \\
Dubna, 141980, Russia\\
E-mail: nestr@thsun1.jinr.ru}

\author{A. FEOLI}

\address{Dipartimento di Ingegneria, Universit\`{a} del Sannio,\\
 Corso Garibaldi n.\ 107, Palazzo Bosco
Lucarelli,\\   Benevento, 82100,  Italy \\
E-mail: feoli@unisannio.it}

\author{G. SCARPETTA}

\address{Dipartimento di Fisica "E.R.Caianiello" --
Universit\`a di Salerno,\\
  Baronissi (SA)   84081, Italy \\
E-mail: scarpetta@sa.infn.it}


\maketitle

\abstracts{The free energy of globular protein chain is considered to be a
functional defined on smooth curves in three dimensional Euclidean space. From
the requirement of geometrical invariance, together with basic facts  on
conformation of helical proteins and dynamical characteristics of the protein
chains, we are able to determine, in a unique way, the exact form of the free
energy functional. Namely, the free energy density should be  a linear function
of the curvature of curves on which the free energy functional is defined. This
model can be used, for example, in Monte Carlo simulations of exhaustive
searching the native stable state of the protein chain.}

\section{Introduction}

A fascinating and open question challenging physics, biochemistry
and even geometry is the presence of highly regular motifs such as
$\alpha-helices$ and $\beta-sheets$ in the folded state of
biopolymers and proteins. A wide range of approaches have been
proposed to rationalize the existence of such secondary structures
(see, for example, reviews\cite{CD,Dill} and references
therein).

We propose a pure geometrical approach\cite{Kam,BM,HAL}
 to describe the free energy of proteins, proceeding from the
most general invariance requirements and basic experimental facts
concerning the protein conformation. Taking into account the
one-dimensional nature of the protein chains, the relevant
macroscopic free energy $F$ should be considered as a functional
defined on smooth curves ${\bf x} (s)$ (or paths) in the three
dimensional Euclidean space
\begin{equation}
\label{eq1}
 F = \int f [{\bf x} (s)]\, d s,
\end{equation}
where $s$ is the length of a protein molecule. The reparametrization invariance
of the functional $F$ demands the free energy density $f$ to be  a scalar
function depending on the geometrical invariants of the position vector ${\bf
x} (s)$, which describes the spatial shape of the protein chain. In three
dimensional ambient space a smooth curve has two local invariants: curvature
$k(s)$ and torsion $\kappa(s)$. In the general case of $D$ dimensional
Euclidean embedding space there are $D - 1$ principal curvatures $k_\alpha (s),
\alpha = 1,2....,D-1$ of a curve, where $k_1 (s) = k(s)$ and $k_2(s) =
\kappa(s)$.

The first principal curvature, or simply the curvature,  $k_1 (s) =
k(s)$ of a curve characterizes the local bending of the curve at the
point $s$. Hence, the dependence of free energy density $f$ on
$k(s)$ specifies the resistance of a protein chain to be bent. The
second curvature or torsion $\kappa(s)$ is determined by the relative
rotation, around the tangent $d{\bf x}(s)/ds$ at the point $s$, of
two neighbor infinitely short elements of the protein chain. It is
well known\cite{CD,Dill} that, in the case of protein molecules, such a
rotation is quite easy, as it requires little effort. In other words,
this rotation results in small energy differences, allowing many
overall conformations of a protein chain to arise. Thus the
dependence of the free energy density $f$ on torsion $\kappa(s)$
can be neglected at least as a first approximation. Finally one can
consider the free energy density $f$ to be a function only of
the curvature $k(s)$, i.e., $ f = f (k(s))$. In what
follows we shall try to specify this dependence explicitly keeping in
mind the description of globular protein conformation.

A peculiarity of conformation of globular proteins is that they can be ordered
assemblies either of helices or of sheets as well as a mixture of helices and
sheets\cite{CD,Dill}. In the phenomenological macroscopic approach, which is developed
here, the presence of sheets in the spatial structure of globular proteins
implies the necessity to introduce, in addition to space curves ${\bf x} (s)$,
new dynamical variables ${\bf y} (s, s')$ describing surfaces in ambient
space.
Obviously such an extension of the problem setting would complicate
considerably our consideration. Therefore we confine ourselves to helical
proteins and try to answer the question: Is it possible to specify the function
$ f (k(s))$ in such a way that the extremals of the functional $ F = \int f\,
ds $
 would be
only helices?  The answer to this question turns out to be positive and unique,
namely, the density of the free energy  $ f (k(s))$ should be a linear function
of the curvature $k(s)$. We sketch here the  proof of this assertion (for
details see Ref.~\cite{arch,NFS1,NFS2}).

\section{Euler-Lagrange equations in terms of principal curvatures
and their exact integrability}
For an arbitrary functional $F$ defined on  curves $x^i (s)$ in
$D$-dimensional space the Euler-Lagrange equations are a set of
exactly $D$ equations
\begin{equation}
\label{eq3}
\frac{\delta F}{\delta x^i} = 0, \quad i= 1,2,\ldots ,D.
 \end{equation}
 However, if the functional $F$ depends
only on the curvature
\begin{equation}
\label{eq4}
F = \int f (k(s))\, ds,
\end{equation}
then $D$ equations (2.3) for
$D$ variables $x^i (s),  \;  i   = 1,2, \ldots  , D$ give $D-1$ equations for
the principal curvatures $k_\alpha (s),  \;  \alpha =  1,2,  \ldots  , D-1$
\begin{eqnarray}
\frac{d^2}{ds^2}\,(f'(k_1)) & =& - \left ( k_1^2-k_2^2\right
)\, f '(k_1)+k_1\,f(k_1)\,{,}\label{eq5}\\
2\frac{d}{ds}\,(f'(k_1)\,k_2) &=&  k^{\prime}_2\,f'(k_1)\,{,}
\label{eq6}\\
k_3 (s) = k_4 (s)= &\ldots& = k_{D-1} (s) = 0.\label{eq7}
\end{eqnarray}

 Thus, in the problem  under consideration there are two non
trivial equations (\ref{eq5}) and (\ref{eq6}) for the curvatures $k_1 (s)$ and
$k_2 (s)$. Equation (\ref{eq6}) can be integrated with arbitrary free
energy density $f (k_1)$
\begin{equation}
\label{eq8}
 \left ( f' (k_1)\right )^2k_2=C\,{,}
\end{equation}
where $C$ is an integration constant.

 Relation (\ref{eq8}) enables one to eliminate the torsion $k_2(s)$
from Eq.\ (\ref{eq5}).
 As
a result we are left with one nonlinear differential equation of
the second order for the curvature $k_1(s)$
\begin{equation}
\label{eq9} \frac{d^2}{ds^2}\,f'(k_1) + \left ( k_1^2  -
\frac{C^2}{(f'(k_1))^4}\right ) f'(k_1) -k_1\,f (k_1) =
0.
\end{equation}

Having resolved this equation for $k_1(s)$, one can determine the rest of
curvatures by making use of Eqs.\ (\ref{eq8})  and (\ref{eq7}). Integration of
the respective Frenet equations with principal curvatures found enables one to
recover the curve $ {\bf x} (s)$ itself.

Notwithstanding its nonlinear character, Eq.\ (\ref{eq9}) can be integrated in
quadratures for arbitrary function $f(k_1)$. To show this, the first integral
for this equation can be constructed proceeding from the symmetry properties of
the variational problem under study\cite{arch}
\begin{equation}
\label{eq10}
M^2 = \left (f'(k_1)\,k_1- f(k_1)\right )^2 +
\frac{C^2}{(f'(k_1))^2}+(k_1')^2(f''(k_1))^2\,{.}
\end{equation}
By direct differentiation of Eq.\ (\ref{eq10}) with respect to $s$ one
can be convinced that, for
\begin{equation}
\label{eq11}
f''(k_1) \not=  0\,{,}
\end{equation}
the relation (\ref{eq10}) is an integral of the nonlinear
differential equation (\ref{eq9}), which determines the
curvature of a stationary curve. From (\ref{eq10}) we deduce
\begin{equation}
\label{eq12}
\frac{dk_1}{ds} = \pm\,\sqrt{g(k_1)}\,{,}
\end{equation}
where
\begin{equation}
\label{eq13}
g(k_1) = \frac{1}{(f''(k_1))^2}\left [ M^2 -\frac{C^2}{(f
'(k_1))^2}-\left ( k_1\,f'(k_1)-f(k_1)\right)^2\right ]\,{.}
\end{equation}
Integration of Eq.\ (\ref{eq12}) gives
\begin{equation}
\label{eq14}
\int_{k_{10}}^{k_1(s)}\!\!\frac{dk}{\sqrt{g(k)}} = \pm\,(s-s_0)
\end{equation}
with $k_{10} = k_1(s_0)$.

Thus, if the free energy density $f (k_1)$ obeys the condition (\ref{eq11}),
then the curvature $k_1 (s)$ and the torsion $k_2 (s)$ of the stationary curve
${\bf x} (s)$ are the functions of the parameter $s$ defined by Eqs.\
(\ref{eq14}) and (\ref{eq8}). The case when the condition (\ref{eq11}) is not
satisfied, i.e., when the free energy density $f (k_1)$ is a linear function of
the curvature $k_1 (s)$, will be considered below.

Now we are going to fix the function $f (k_1)$, requiring that all the
solutions to the Euler - Lagrange equations are helices. From the differential
geometry of curves it is known that the helices in three
dimensional space have a constant curvature $(k_1)$ and a constant torsion
$(k_2)$ which determine the radius $R$ and the step $d$ of a helix
\begin{equation}
\label{eq15}
R = \frac{k_1}{k^{2}_1 + k^{2}_2}, \quad d = \frac{2\pi
|k_2|}{k^{2}_1 + k^{2}_2}.
\end{equation}
Under the condition (\ref{eq11}) the curvature $k_1$ and the torsion $k_2$ of
the stationary curve are not constant but they are the functions of the
parameter $s$ which are defined in Eqs.\ (\ref{eq14}) and (\ref{eq8}). Hence,
for the free energy density $f (k_1)$ we are looking for, we have to consider
the case, when
\begin{equation}
\label{eq16}
f'' (k_1) = 0,
\end{equation}
i.e.,  $f (k_1)$ is  a linear function of the curvature $k_1 (s)$
\begin{equation}
\label{eq17}
f (k_1) = \alpha + \beta \, k_1(s),
\end{equation}
where $\alpha$ and $\beta$ are constants. Indeed, substituting Eq.\
(\ref{eq17}) into (\ref{eq8}) and (\ref{eq9}) we obtain the constant curvatures
$k_1$ and $k_2$:
\begin{equation}
\label{eq18}
k_1 = - \frac{C^2}{\alpha  \beta^3}, \quad  k_2 =
\frac{C}{\beta^2}\,{.}
\end{equation}
Since $k_1 (s) = |{\bf x}'' (s)| > 0$, the constants $\alpha$ and
$\beta$ should have opposite signs. It is natural to put $\alpha > 0$
and $\beta < 0$.

For the free energy density $f(k_1)$, linear in curvature, the integral
(\ref{eq10}) gives just the relation between the integration constants $M^2$
and $C^2$: $ M^2[1-(C/\beta M)^2] =  \alpha^2$.

When $\alpha = 0$, Eq.\ (\ref{eq9}) implies that the curvature $k_1 (s)$
is an arbitrary function of $s$ and the integration constant $C$
vanishes. In this case Eq.\ (\ref{eq8}) yields $k_2 = 0$. Hence, for
$\alpha = 0$, the solutions to the Euler-Lagrange equations are
arbitrary plane curves, which is evidently unacceptable for our
purpose.

 Finally, requiring that the stationary curves for the
functional (\ref{eq4}) are only helices, we uniquely determine  the free
energy density, namely it should be a linear function of the
curvature
\begin{equation}
\label{eq20}
f (k_1) = \alpha - |\beta|\, k_1(s)
\end{equation}
with nonzero constants $\alpha$ and $\beta$, providing $\alpha > 0$
and $\beta < 0$. From the mathematical stand point, we have resolved in fact
the inversion variational problem, i.e., proceeding from the specified solutions
to the variational Euler-Lagrange equations (helical curves) we have uniquely
recovered the respective functional.

Closing this section we would like to make the following note. It is well
known, that in the protein physics the chirality property of these  molecules
is important. At the first glance we have ignored this point because from the
very beginning we have eliminated the dependence of the free energy density $f$
on the curve torsion. But the fact is the integration constant $C$ in Eq.\
(\ref{eq8}) is responsible for the chirality property of the helical curves
under consideration. Really, this constant is an analog of the Pauli-Lubanski
pseudoscalar\cite{arch,N1,N2} (in the three dimensional space-time the
Pauli-Lubanski vector reduces to the pseudoscalar). The positive and negative
values of this constant distinguish the left-hand and right-hand chirality of
the helical curves.

\section {Conclusion}
Proceeding from rather general principles and making use of the basic facts
concerning the conformation of globular proteins we have obtained, in a unique
way, a geometrical model for phenomenological description of the free energy of
helical proteins. It is worth noting that our functional (\ref{eq20}) should be
considered as an effective  free energy of the helical protein which already
takes into account the n atomic interactions within the protein and with the
solvent. Hence, there is no need to quantize it, as one proceeds in the random
walk studies.\cite{r1,r2,r3}

Certainly our simple model does not pretend to describe all the aspects of the
protein physics. However, one can hope that it could be employed, for example,
in Monte Carlo simulation to search for a stable native state of the protein.
In this case the model can be used for the description of the free energy of
individual parts (blocks) of a protein chain that have the helical form.
Without any doubt, it should result in simplification and acceleration of the
exhausting searching of the native stable state of the protein chain by a
computer.\cite{CD,Dill}


\begin{thebibliography}{99}
\bibitem{CD} H.\ S.\ Chan  and K.\ A.\ Dill, {\it Physics Today}
{\bf 46},  No.\ 2, 24  (1993).

\bibitem{Dill} K.\ A.\ Dill,  {\it Protein Science} {\bf 8},
  1166 (1999).

\bibitem{Kam} R.\ Kamien, {\it Rev. Mod. Phys.} {\bf 74}, 954 (2003).

\bibitem{BM} J.\ R.\ Banavar and A. Maritan,
{\it Rev. Mod. Phys.} {\bf 75}, 23 (2003).

\bibitem{HAL} S.\ Hyde, S.\ Anderson, K.\ Larsson. {\it The Language of Shape}.
Elsevier, Amsterdam, 1997.

\bibitem{arch}
 A.\ Feoli, V.\ V.\ Nesterenko, and G.\ Scarpetta, cond-mat/0211415.

\bibitem{NFS1} V. V. Nesterenko, A. Feoli and G. Scarpetta,
{\it J.\ Math.\ Phys.} {\bf 36}, 5552 (1995).

\bibitem{NFS2} V. V. Nesterenko, A. Feoli and G. Scarpetta,
 {\it Class.\ Quantum Grav.}  {\bf 13}, 1201 (1996).

\bibitem{r1}
R. D.\ Pisarski, {\it Phys.\ Rev.}  {\bf D34}, 670 (1986).

\bibitem{r2}
J.\ Ambj\"orn, B.\ Durhuus, and T.\ Jonsson, {\it J.\ Phys.}  {\bf A21},
981 (1988).

\bibitem{r3}
 A.\ L.\ Kholodenko,  {\it Ann.\ Phys.}  {\bf 202}, 186 (1990).

\bibitem{N1} V.\ V.\ Nesterenko, {\it J.\ Math.\ Phys.} {\bf 32}, 3315 (1991).

\bibitem{N2}  V.\ V.\ Nesterenko, {\it J.\ Math.\ Phys.} {\bf 34}, 5589 (1993).

\end{thebibliography}
\end{document}